\documentclass[twocolumn, aps, amssymb, prl ]{revtex4-1} 
\usepackage{amsmath, amssymb, amsthm, wasysym}

\usepackage[section]{placeins}

\usepackage{rotating}
\usepackage{floatpag}
\rotfloatpagestyle{empty}
\usepackage{graphicx}
\usepackage{multind}
\usepackage{hyperref}
\usepackage{epstopdf}

\begin{document}

\title{Light propagation and emission in complex photonic media}

\author{Willem L. Vos, Ad Lagendijk, and Allard P. Mosk}
\affiliation{Complex Photonic Systems (COPS), MESA+ Institute for Nanotechnology, University of Twente, P.O. Box 217, 7500 AE Enschede, The Netherlands}

\date{Prepared on May 21st, 2013, corrected October 26th, 2014. \\ .. From: \emph{Light Localisation and Lasing}, Eds. M. Ghulinyan, L. Pavesi, Cambridge Univ. Press (2015) Ch. 1, p. 1}

\begin{abstract}
ABSTRACT 
We provide an introduction to complex photonic media, that is, composite materials with spatial inhomogeneities that are distributed over length scales comparable to or smaller than the wavelength of light. 
This blossoming field is firmly rooted in condensed matter physics, in optics, and in materials science. 
Many stimulating analogies exist with other wave phenomena such as sound and seismology, X-rays, neutrons. 
The field has a rich history, which has led to many applications in lighting, novel lasers, light harvesting, microscopy, and bio optics. 
We provide a brief overview of complex photonic media with different classes of spatial order, varying from completely random to long-periodically ordered structures, quasi crystalline and aperiodic structures, and arrays of cavities. 
In addition to shaping optical waves by suitable photonic nanostructures, the realization is quickly arising that the spatial shaping of optical wavefronts with spatial light modulators dramatically increases the number of control parameters. 
As a result, it is becoming possible for instance to literally see through completely opaque complex media. 
We discuss a unified view of complex photonic media by means of a photonic interaction strength parameter.
This parameter gauges the interaction of light with any complex photonic medium, and allows to compare complex media from different classes for similar applications. 
\\

\end{abstract}
\maketitle

\section*{General overview}
In many areas in the physical sciences, the propagation of waves in complex media plays a central role. 
Acoustics, applied mathematics, elastics, environmental sciences, mechanics, marine sciences, medical sciences, microwaves, and seismology are just a few examples, and of course nanophotonics~\cite{Sheng1995, Soukoulis1996, Soukoulis2001, Akkermans2007}. 
Here, complex media are understood to exhibit a strongly inhomogeneous spatial structure, which determines to a large extent their linear and nonlinear properties. 
Examples of complex media are random, heterogeneous, porous, and fractal media.
The challenges that researchers in these areas must surmount to describe, understand, and ultimately predict wave propagation are formidable. 
Right from the start of this field - in the early 1900s - it was clear that new concepts and major approximations had to be introduced. 
Famous examples of such concepts are the effective medium theory~\cite{MaxwellGarnett1904PTRS, Bruggeman1935AP, Bohren1983} and the radiative transport theory~\cite{Chandrasekhar1950, vandeHulst1957}. 
These concepts are very much alive, even today. 
The fundamental challenge with these approximate concepts is that often the length scales of the inhomogeneities in the complex medium are comparable to the wavelength, whereas the range of validity of these approximations is restricted to situations where these length scales are much larger than the wavelength. 
Consequently, many relevant situations arise where either the effective medium theory, or radiative transport theory, or both, fail dramatically. 
Examples of such situations are given in this introduction and throughout this book. 

The complexity of the medium that supports wave propagation can be classified in a number of ways. 
Many different types of spatial inhomogeneities in a host matrix can be envisioned, varying from completely random, via aperiodic and waveguide structures to quasi-crystalline and even structures with long-range periodic order. 
The inhomogeneities could be of a self-organized form, or the fruit of precise engineering. 
An additional classification is whether or not the complexity is confined to the surface (in two dimensions (2D)) or is present throughout the volume of the medium (in three dimensions (3D)), as in porous media. 
A useful criterion for the classication of inhomogeneous media is whether the inhomogeneity is of a continuous nature, or stems from discrete scatterers. 
One can thus classify the topology of the inhomogeneities relevant to classical waves: if the inhomogeneities are connected from one side to the other, the medium has a network topology; if the inhomogeneities are completely surrounded by host material, the complex medium is said to have a Cermet topology. 
It appears that topology plays an important role in the scattering of various types of waves - such as scalar, elastic, electromagnetic - in complex media~\cite{Economou1993PRB}. 

Ever since the breakthrough achievement of renormalization group theory~\cite{Wilson1971PRB} the dimensionality of a complex system is known to be vital~\cite{Chaikin2000, Sheng1995}. 
Hence, the dimensionality of the problem is crucial to the study of waves in complex media. 
1D and 2D systems have the intriguing property that waves are always localized, that is, a wave that starts at a certain spatial position always returns. 
As stated by the Mermin-–Wagner–-Hohenberg theorem, 2D systems have the lower critical dimension for most field theories~\cite{Mermin1966PRL, Hohenberg1967PR}, allowing for the study and use of a wealth of scaling phenomena that are highly challenging to uncover.
In a 3D world there is a striking phase transition between localized and extended states. 
Famous examples are Anderson localization, or the photonic band gap in 3D. 

From scaling theory~\cite{Abraham1979PRL, Sheng1995}, it is known the extent of the complex system is crucial. 
The finite size determines the transport of the waves - specifically, the conductance.
For extended states, it appears that the conductance scales with dimensionality minus two, times the logarithm of the system size.
For localized states, the conductance decreases exponentially with system size. 
Therefore, the study of system size dependence of complex media provides an important key to distinguish extended from localized states, and to characterize the formation of gaps. 

Advances made in the understanding of waves in complex media have led to a number of practical applications, and are generating new ones at an ever increasing pace~\cite{Wiersma2013NP}. 
Examples are found in remote optical sensing ("looking through a cloud"), inverse optical scattering, noninvasive medical imaging ("find the tumour in tissue"), applied optics (quality control of optical systems by controlling surface roughness), optical devices and (random) lasers, furthermore in oil and mineral prospecting by seismic methods, in ultrasonic imaging and non-destructive testing ("find hairline cracks in an airplane wing"), material characterization, all the way to microwave propagation and detection in antennas, mobile phones, and radar. 
Researchers working with waves in complex media can be thus found in many areas of the engineering sciences, physics, and chemistry. 
Concepts developed in one of the individual disciplines have found or are bound to find uses in other wave disciplines. 
It has thus become increasingly clear that multidisciplinary contacts are very fruitful. 
Important experimental discoveries and theoretical breakthroughs made in one of these individual wave disciplines frequently appear to have important consequences for all the other wave disciplines. 

These rapid developments and cross-fertilization have led to an enormous increase in the understanding of wave propagation that has meanwhile advanced to a very high level~\cite{Beenakker1997RMP, Akkermans2007}.
As a result, we are reaching the point where we can incorporate complex wave behavior in applied techniques as lidar, sonar, and radar. 
The fast rise of the field of medical imaging with diffuse waves is also stimulated by this tremendous progress.
In all cases, the advances pertain to new concepts, to new experimental discoveries, and to new techniques.

\section*{Light in complex photonic media}
The exciting subject of light in complex photonic media has become a very active area of research and grown to become a field of its own. 
The field has its roots in the physics of electrons and spins in condensed matter~\cite{Anderson1958PR, Sheng1995, vanRossum1999RMP, Akkermans2007}. 
The essential optical properties of complex photonic media are determined by the spatially varying refractive index, analogous to the periodic potential for an electron in condensed matter. 
Large variations of the refractive index cause a strong interaction between light and the photonic medium.
As a result, multiple scattering dominates the linear and non-linear optical properties. 
As complex photonic media, we consider in this Introductory chapter random media with short-range order, photonic crystals with long-range periodic order, arrays of cavities or waveguides, quasi crystals with orientational order yet without long-range order, or even aperiodic media. 
In this section we give a brief overview of these classes of complex media in sequence of their historical appearance, and the main research directions as illustrated with a sample of key references. 

\begin{figure}
\includegraphics[width=1.0\columnwidth]{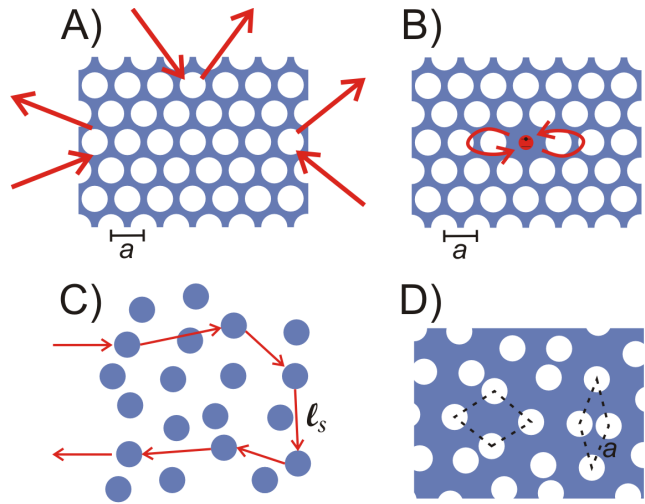}
\caption{
Schematic cross-sections of the main nanophotonic complex media discussed in this chapter.
A) A photonic band gap crystal: incident light within a particular frequency band - shown as red arrows - is multiply diffracted by all crystal planes, irrespective of the direction of incidence. 
Hence, it is dark inside the crystal in the band gap as a result of interference.
B) A cavity in a photonic band gap crystal. 
C) A random medium with mean free path $\ell_{s}$ step size between scattering events. 
D) A quasi crystal, Penrose tiling. 
The two kinds of parallellograms ("fat" and "skinny") are indicated by dashed lines. 
The lattice parameter $a$ is equal to the edge length of the constituent parallelograms. 
}
\label{fig:classes-of-complex-media}
\end{figure}

\subsection{Random media}

The study of light propagation in random media is a field of research with a rich history starting in the 1980s~\cite{Ishimaru1978, vandeHulst1957}. 
In daily life, we encounter light in disordered dielectric media, such as paint, milk, fog, clouds, or biological tissue.
Light performs a random walk through the complex medium, as illustrated in Fig.~\ref{fig:classes-of-complex-media}(C). 
The average step size in the random walk is equal to the scattering mean free path $\ell_{s}$. 
In the above mentioned media, the mean free path is (much) larger than the wavelength, $\ell_{s} > \lambda$, and the transport of light is well described by diffusion, as if light scatters like particles~\cite{vanRossum1999RMP, Akkermans2007}. 

When the average distance between scattering events is reduced to become comparable to the wavelength of light $\ell_{s} \simeq \lambda$, interference cannot be neglected anymore. 
This realization came as a shock, since it was originally thought that random multiple scattering scrambles all phases. 
Experimentally, the signature of interference in complex media is observed as an enhanced backscatter cone in the intensity - even for relatively weakly interacting samples. 
The cone is the result of constructive interference between light that propagates along any random path in the complex medium, and light that propagates along the time-reversed path~\cite{vanAlbada1985PRL, Wolf1985PRL}. 
The width of the cone is inversely proportional to the mean free path, and thus serves as a convenient probe thereof. 
The mean free path decreases when the interaction between light and the complex medium increases.  

Ultimately, interference causes transport to grind to a complete halt, a phenomenon known as Anderson localization of light~\cite{Anderson1958PR, John1984PRL, Lagendijk2009PT}. 
The diffusion of light tends to zero at the Anderson localization transition, as light inside the medium has a very high probability to return to its original position. 
While many efforts have been devoted to the observation of Anderson localization of light in 3D, there have been only few reports to date, using strongly scattering powders of GaAs~\cite{Wiersma1997N} and titania~\cite{Storzer2006PRL, Sperling2013NP}. 
Currently, the consensus seems to be that the observations are compounded by optical absorption, nonlinear effects, or spurious fluorescence~\cite{VanderBeek2012PRB, Wiersma2013NP}. 
Therefore, the quest for Anderson localization of light at optical frequencies in 3D is still open. 
Experimental techniques that are likely to settle debates on localization versus spurious effects is the probing of optical correlations, as demonstrated for microwaves~\cite{Chabanov2000N}, see the chapter by Genack and Shi. 
As we have seen above, Anderson localization is easier to observe in lower dimensions. 
This has lead to a renewed interest in this striking wave phenomenon with the advent of quasicrystals and aperiodic media, as well as in waveguides and coupled cavity arrays, see the next subsections. 

Optical waves that are transmitted through or reflected from a complex medium also reveal another interference phenomenon, namely speckle. 
By eye, speckle is observed as the grainy pattern in a light spot on a wall that is illuminated with a laser pointer. 
Speckle is a random interference pattern that depends very sensitively on the detailed arrangement of all scatterers in the complex medium. 
Interestingly, the realization has recently arisen that the speckle pattern can be drastically modified by playing on the spatial wavefronts of the incident waves, see the next section. 
The angular distribution of light in a speckle pattern is impossible to predict, a property that can be applied to apply random media as physically unclonable objects for encryption~\cite{Pappu2002S}. 
By averaging over the positions of the scatterers, or by averaging over the optical frequencies, the interference washes out and a speckle pattern reverts to a diffuse distribution. 
Nevertheless, speckle has well-defined statistical properties (intensity distribution).
Moreover speckle contains three types of correlations that depend sensitively on the light propagation inside the complex medium~\cite{vanRossum1999RMP}. 
First, if the frequency of the incident light is changed, the speckle pattern deforms as described by the short-ranged $C_{1}$ correlation function. 
Likewise, if the direction of the incident beam is changed, the speckle pattern deforms as again described by the $C_{1}$ correlation function. 
Simultaneously, the speckle pattern follows the incoming beam, known as the memory effect. 
Secondly, long-ranged correlations occur even between different speckle spots, as described by the $C_{2}$ correlation function. 
Thirdly, infinite-range correlations occur when one averages over all incident and outgoing directions. 
These so-called $C_{3}$ correlations are also known as universal conductance fluctuations, and have been observed in beautiful experiments by Scheffold and Maret~\cite{Scheffold1998PRL}. 

Separate from propagation and transport effects as described above, random complex media are also pursued for their effects on the local density of optical states (LDOS) that notably controls the emission of an embedded light source. 
Inspired by work on photonic crystals - in themselves an offspring of random media - it is being realized that random media may also open a 3D photonic bandgap. 
Calculations supported by microwave experiments predict that a 3D photonic bandgap will occur in amorphous diamond~\cite{Imagawa2010PRB, Liew2011PRA}. 
This pursuit is also inspired by condensed matter, where it is well-known that long-range periodic order is not crucial to sustain a bandgap, as both crystalline and amorphous silicon possess such a gap. 

In random media, it has been predicted that the LDOS displays intricate spatial fluctuations that are described by the $C_{0}$ correlation function~\citep{Shapiro1999PRL}. 
As opposed to the correlation functions $C_{1}, C_{2}, C_{3}$ above that have analogies in transport of electrons, $C_{0}$ is peculiar to photonic media, as spontaneous emission is peculiar to photonic media. 
The first intriguing effects have recently been observed by studying light emitters inside random photonic media~\cite{Birowosuto2010PRL, Sapienza2011PRL}, on the surface of random media~\cite{Ruijgrok2010OE}, and even inside random plasmonic media~\cite{Krachmalnicoff2010PRL}. 

Whereas in a conventional laser mirrors provide feedback, a random laser uses multiple scattering of light as a feedback mechanism. 
In a random laser, a spontaneously emitted photon is amplified by stimulated emission. 
Simultaneously its path length in the gain medium is increased. 
Consequently the emission spectrum narrows for increasing pump powers and the output power for the peak
of the spectrum shows typical threshold behavior. 
In contrast to a conventional laser the emitted light from a random laser is omnidirectional.
The idea of generating light inside a scattering medium by stimulated emission was proposed in 1968 by Letokhov~\cite{Letokhov1968JETP}. 
After initial work, the field took off in the mid 1990s fueled by a debate whether laser action in scattering samples was truly due to diffusion or to single scattering~\cite{Lawandy1994N, Wiersma1995N}. 
The involved Nature editor coined the term "random laser" which has remained \emph{en vogue} ever since. 
A random laser can be conceived as a multiple scattering medium with gain or as a laser system with a complex cavity configuration. 
Well-known laser physics effects are observed with random lasers, such as relaxation oscillations, intensity fluctuations, mode coupling, as well as prominent multiple scattering phenomena such as enhanced backscattering and speckle, see the reviews~\cite{Cao2005JPA, Wiersma2008NP}. 

In pioneering studies Cao and co-workers detected narrow features in the output spectrum of random lasers, called 'spikes'~\cite{Cao1999PRL}. 
Interestingly the community has not yet converged on an interpretation of these intriguing narrow spectral features. 
While impressive theoretical efforts have focussed on 2D random laser systems~\cite{Tureci2008S}, it remains to be seen how the theoretical concepts translate to 3D systems and how these theories can be connected to experiments. 
Anderson localization, Fabry-Perot resonances, absorption induced confinement, and photons traveling exceptionally far through the gain medium, are examples of explanations put forward that illustrate the many proposed interpretations, see~\cite{Cao2005JPA, Wiersma2008NP}. 
We recommend the chapter by Leonetti and Lopez for further review on this exciting subject.

\subsection{Photonic band gap crystals}
Photonic crystals have become a subject of intense research in the late 1990s and early 2000s~\cite{Soukoulis1996, Soukoulis2001}. 
We will defer the discussion of 3D bandgap-induced cavity QED phenomena such as spontaneous emission inhibition, Purcell-enhanced emission, or thresholdless laser action, to the relevant chapter by Vos and Woldering in this book \footnote{See http://cops.nano-cops.com/publications/cavity-quantum-electrodynamics-three-dimensional-photonic-bandgap-crystals. This is an extra reference compared to the published version.}. 

The field of photonic band gap crystals is intimately linked to that of random media by the role of interference in modifying the transport of light. 
Indeed, one of the original drives for photonic band gaps was the prediction that Anderson localization are more easily reached in photonic band gap crystals with controlled disorder~\cite{John1987PRL}. 
Moreover, photonic crystals are being pursued for the flexible possibilities to control the propagation of light, ranging from simple filter action in reflectivity or transmission, to superprisms and negative refraction, to ultimately confined yet broadband waveguiding in 3D, see ~\cite{Lourtioz2008}.

\subsection{Waveguides and coupled cavities}
One-dimensional (1D) waveguides in photonic crystals - mostly in 2D slab crystals - and arrays of cavities have received great interest since the mid 1990s, motivated by applications of slow light, and fueled by on-chip optical communication~\cite{Krauss2007JPD, Baba2008NP}. 
For a long time, the effects of unavoidable variations in size and position of the building blocks on the propagation of light were neglected~\cite{Koenderink2005PRB}. 
Ever since, localization has received revived interest in the second half of the 2000s.
The interest arose in the context of randomness in coupled cavities, and randomness in photonic crystal waveguides~\cite{Topolancik2007PRL, Mookherjea2008NP}. 
Recently, it has even been realized that in presence of 1D Anderson localization in waveguides, spontaneous emission of embedded quantum dots is considerably enhanced~\cite{Sapienza2010S}. 
An even stronger analogy between the behavior of light and of electrons in condensed matter physics is the recent observation of Lifshitz tails in photonic crystal waveguides~\cite{Huisman2012PRB}. 
For a review of these exciting phenomena, see the chapter by Mookherjea in this book.

\subsection{Quasicrystals and aperiodic media}
Quasicrystals and aperiodic photonic media have received attention since the 2000s. 
The pursuit of quasicrystals has been motivated by the notion to open photonic gaps at low index contrast, thereby allowing a wide choice of materials, including biomaterials. 
This motivation was fueled by reports where a 2D gap was found to open at a low index contrast $m \geq 1.45$~\cite{Zoorob2000N}. 
Subsequent work revealed that a gap opens for $m \geq 2.6$~\cite{Zhang2001PRB}. 
Nevertheless, 2D quasicrystals are generating exciting results and are being pursued for their strong light scattering properties and intricate random laser action.
In addition deterministic and multifrequency laser emission is being reported from aperiodic media~\cite{DalNegro2012LPR}, see the chapter by Cao, Noh, and Dal Negro. 

In the case of 1D quasicrystalline structures, pioneering work was reported by Ref.~\cite{DalNegro2003PRL}, where intricate 
pulse stretching was observed, as well as a strongly suppressed group velocity for frequencies close to a Fibonacci band gap.
For more review on this exciting subject, see the chapter by Ghulinyan.  

In the case of 3D quasicrystals initial studies were reported in the microwave regime, where crystals with different structures can conveniently be machined~\cite{Man2005N}. 
Work at optical frequencies was reported by Ref.~\cite{Ledermann2006NM}, where the required nanostructures were made with advanced laser writing technies. 
By studying the transport of light with ultrashort pulses, a very short mean free path was observed which points to incipient localization effects in these intriguing structures~\cite{Ledermann2006NM}. 
For an in-depth review see the chapter by Ledermann, Renner, and von Freymann.

\section*{Shaping wave fronts in complex media}
In complex media such as photonic crystals, light is strongly scattered both at the interfaces and in the bulk of the material. Due to unavoidable disorder in the structure, random scattering adds to the desired diffraction~\cite{Koenderink2005PRB}. 
As a result, it is impossible to directly image or focus light deep inside strongly photonic nanosctructures, and excitation and light collection from sources inside photonic crystals is typically inefficient.

\begin{figure}
\includegraphics[width=1.0\columnwidth]{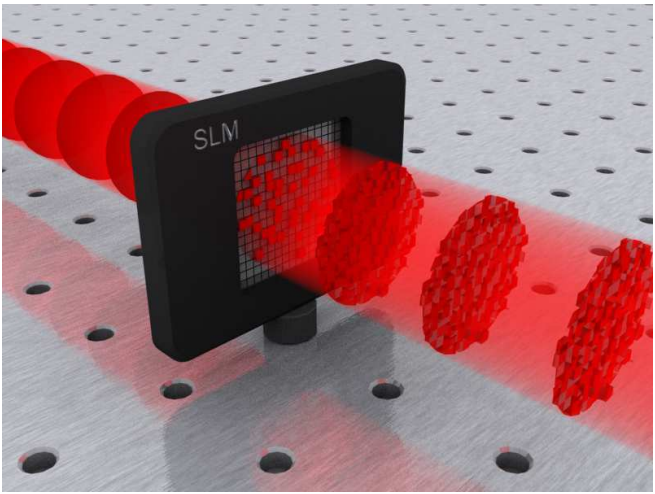}
\caption{
Schematic image of incident plane waves from the left. 
In a spatial light modulator (SLM) the plane wave fronts are modified into complex phase patterns (right) that propagate to a complex medium. 
}
\label{fig:SLM-for-WFS}
\end{figure}

Recently developed wavefront shaping methods, which are leading to a boost of activity in optics of complex media, offer a method to counteract this unavoidable disorder, and even exploit it to obtain a high focusing resolution~\cite{Mosk2012NP}. 
These methods make use of the fact that scattered light does not lose its coherence, so that the incident waves can be forced to interfere constructively at a chosen point, giving rise to a focus of high intensity. 
With the help of megapixel spatial light modulators, see Fig.~\ref{fig:SLM-for-WFS}, optical wavefronts are spatially shaped and adapted with sufficient precision to counteract and even exploit scattering. 
Already in the very first wavefront shaping experiments, light was focused through opaque media that were more than 15 mean free paths thick and it was shown that a focus can be created that is 1000 times more intense than the diffusive speckle background~\cite{Vellekoop2007OL}. 
Using fluorescent or nonlinear probes, light can also be focused inside a scattering medium~\cite{Vellekoop2008OE}. 
A particularly beneficial feature in emission experiments is that due to reciprocity a phase screen that optimizes excitation will also optimize light collection, provided that the processes take place at near enough wavelengths. 
By scanning the optimized spot, imaging with limited range is possible in several configurations~\cite{Vellekoop2010OL, Hsieh2010OE}. 
An important step towards even more flexible focusing and imaging was taken by Popoff and coworkers, who used wavefront shaping methods to measure a transmission matrix of a disordered nanophotonic system~\cite{Popoff2010PRL}. 
With the help of such a transmission matrix one can image through scattering media without scanning and focus light at a chosen point in the transmission plane without further optimization. 

Random nanophotonic materials excert strong control over light, e.g. scattering it to high transversal wave vectors that are not accessible by normal refraction. 
Using wavefront shaping, this scattered light can be focused to a small spot and used for imaging at a resolution better than 100 nm~\cite{vanPutten2011PRL}. 
The combination of wavefront shaping with specially designed plasmonic materials is expected to give rise to an even higher imaging resolution~\cite{Lemoult2010PRL, Gjonaj2011NP, Kao2011PRL}. 
Developments in imaging using shaped wavefronts are progressing quickly. 
Recent work has enabled see-through imaging based on speckle correlations\cite{Bertolotti2012N}. 
In this imaging modality, no calibration or probing in the object space is necessary, and the resolution of the image is essentially set by the diffraction limit.

While most of the work so far has involved narrowband light, recent progress in pulse shaping in disordered materials has shown that sufficient bandwidth is available to focus, delay and compress femtosecond pulses through optically thick media~\cite{Aulbach2011PRL, McCabe2011NC, Katz2011NP}. 
This tailoring of broadband pulses is of special importance in photonic media the cases of nonlinear excitation and ultrafast optical switching.

Contrary to the situation in biomedical settings, in nanophotonics one often deals with time-invariant structures and photostable emitters. 
This makes it possible to perform accurate optimizations of even small signals, as one can extend the integration time to obtain good signal to noise. 
This should enable wavefront-based optimization of entirely new signals such as the weak spontaneous emission from emitters deep inside a photonic bandgap crystal.

\section*{Unified view of complex media: photonic interaction strength}
While the various classes of complex media reviewed above have very different order, we propose here a unified view to characterize light propagation in all these different kinds of complex media. 
To this end we define a figure of merit that gauges the interaction with light for all possible complex media.
Such a figure of merit allows to compare examples from the different classes of complex media when they are considered for a similar application. 
The photonic interaction strength $S$ is defined as the dimensionless ratio of the polarizability $\alpha$ of an average scatterer in a complex medium to the average volume per scatterer $V$~\cite{Vos1996PRB, Vos2001Kre, Koenderink2003the}: 
\begin{equation}\label{eq:S-definition}
S = \frac{4 \pi \alpha}{V},
\end{equation}
where the prefactor depends on the chosen electrodynamic units. 
The photonic interaction strength can be interpreted as the ratio of the "optical volume", i.e., the volume that light "experiences" at each scatterer, to the physical volume of each scatterer. 
The feature that these two volumes are not the same can be readily appreciated from the well-known fact that an atom on resonance - such as alkali atom near the D-lines - has a polarizability $\alpha \simeq \lambda^3 \simeq 1~\mu$m$^3$, which is much greater than the physical volume of an atom of about ($0.3$ nm)$^3$ = $3 \times 10^{-11}~\mu$m$^3$. 
Based on considerations for periodic systems~\cite{Koenderink2003the}, one can rewrite the photonic strength as 
\begin{equation}\label{eq:S-recip-space}
S = \frac{|\Delta \epsilon|}{\overline{\epsilon}} |f(\Delta \mathbf{k})|,
\end{equation}
where $\Delta \epsilon$ is the spatial modulation of the dielectric function in the complex medium, $\overline{\epsilon}$ is the volume-averaged dielectric function, and $f(\Delta \mathbf{k})$ is the medium's structure factor evaluated at a dominant scattering vector $\Delta \mathbf{k}$~\citep{Chaikin2000}. 
For dielectric media, Eq.~(\ref{eq:S-recip-space}) points to three main ways to increase the photonic strength. 
First, increasing the dielectric contrast $\Delta \epsilon$, which rationalizes the pursuit of materials with a high refractive index contrast such as semiconductors Si, GaP, GaAs. 
Secondly, decreasing the average dielectric constant $\overline{\epsilon}$, which explains the prevalence of a minority of high-index material in strongly interacting photonic media. 
Thirdly, optimizing the geometrical factor $f(\Delta \mathbf{k})$, which directs choices to specific structural geometries such as the diamond structure for 3D photonic crystals~\cite{Ho1990PRL}, or which motivates the pursuit of quasi-crystals~\cite{Man2005N}. 

Let us provide typical magnitudes of $S$ for well-known examples of complex media. 
For Anderson localization and photonic gap behavior at optical frequencies the strength must be in the range $S > 0.15$ to $0.20$. 
Atoms in a crystal lattice scatter X-rays only very weakly since the refractive index for X-rays hardly differs from 1, hence the photonic interaction strength is very small $S \approx 10^{-4}$.
This result explains why the vast literature on X-ray scattering does not consider Anderson localization or photonic bandgaps for X-rays. 
A widely studied optical material such as an opal made from silica or polystyrene nanospheres has a moderate photonic interaction strength near $S = 0.06$, on account of the moderate refractive index contrast. 
Complex photonic media made from high refractive index semiconductors have high photonic strengths of more than $0.2$. 
Among the highest photonic strengths are the ones in a narrow frequency range about an atomic resonance, for Cs atoms near $\lambda = 850$ nm the strength is about $S = 1.0$. 

Interestingly, the photonic interaction strength is related to the characteristic length optical scale relevant to the particular complex medium under study. 
For random photonic media, the characteristic length optical scale is the scattering mean free path $\ell_{sc}$, the average step size in the random walk of light. 
In photonic crystals with long-range periodic order, the characteristic length optical scale is the Bragg attenuation length $\ell_{Bragg}$, that is, the distance needed to build up Bragg interference such that transmission in a gap has decreased to $1/e$. 
Indeed, the characteristic length optical scale $\ell_{X}$ ($X$ = sc, Bragg, etc.) is inversely related to the photonic interaction strength by 
\begin{equation}\label{eq:ell-vs-S}
\ell_{X} = \frac{\lambda}{\pi S}.
\end{equation}
This relation shows that a strongly interacting complex photonic medium has a short characteristic length scale $\ell_{X}$. 
For instance, in a strongly interacting random medium it takes only a short distance to develop the optical interference required for Anderson localization. 
Since any real complex photonic medium is necessarily finite, it follows that for a complex medium to reveal the desired characteristics (e.g. localization, bandgaps), its extent $L$ should exceed the characteristic length scale: $L > \ell_{X}$.


\section*{Acknowledgments}
We thank L\'eon Woldering and Florian Sterl for help with the illustrations. 
It is a great pleasure to thank all our colleagues at COPS for many years of pleasant and fruitful collaborations, including Cock Harteveld, Femius Koenderink, Peter Lodahl, Pepijn Pinkse, all our postdocs, PhD students and undergraduate students, and many, many others. 
We also thank colleagues elsewhere such as Sergey Skipetrov, Bart van Tiggelen, Diederik Wiersma, and many many others. 
This work is part of the research program of the ``Stichting voor Technische Wetenschappen (STW),'' and the ``Stichting voor Fundamenteel Onderzoek der Materie (FOM),'' which are financially supported by the ``Nederlandse Organisatie voor Wetenschappelijk Onderzoek (NWO)'', and is also supported by the ERC.

\bibliographystyle{cambridgeauthordate}
\bibliography{Complex-photonic-media_Intro_Vos}\label{refs}

\end{document}